\documentclass{eas}
\usepackage{graphicx}
\begin{document}

\title{The variable Universe through the eyes of Gaia} 
\author{Laurent Eyer}\address{Observatoire de Gen\`{e}ve, Universit\'{e} de Gen\`{e}ve, CH-1290 Sauverny, Switzerland}
\author{Maria Suveges}\address{ISDC, Observatoire de Gen\`{e}ve, Universit\'{e} de Gen\`{e}ve, CH-1290 Sauverny, Switzerland}
\author{Pierre Dubath$^2$}
\author{Nami Mowlavi$^2$}
\author{Claudia Greco$^1$}
\author{Mihaly Varadi$^1$}
\author{Dafydd W. Evans}\address{Institute of Astronomy, University of Cambridge, Cambridge CB3 0HA, UK}
\author{Paul Bartholdi$^1$}
\begin{abstract}
The ESA Gaia mission will provide a multi-epoch database for a billion of objects, including variable objects that comprise stars, active galactic nuclei and asteroids. We highlight a few of Gaia's properties that will benefit the study of variable objects, and illustrate with two examples the work being done in the preparation of the data processing and object characterization.
The first example relates to the analysis of the nearly simultaneous multi-band data of Gaia with the Principal Component Analysis techniques, and the second example concerns the classification of Gaia time series into variability types.
The results of the ground-based processing of Gaia's variable objects data will be made available to the scientific community through the intermediate and final ESA releases throughout the mission. 

%

\end{abstract}
\maketitle
\section{Introduction}
When we think about the Universe, we are overwhelmed not only by its hugeness, but also by its diversity. One aspect of this diversity expresses itself in the time domain. Long thought to be immutable, many objects in the Cosmos show variabilities that are observable on a human time-scale. Indeed, the variability properties of observed light curves range from very regular pulsations like Cepheids to unique events such as supernovae explosions. They also cover time scales from the order of milliseconds for $\gamma$-Ray Bursts to centuries for stellar secular evolution, and show amplitude variations from parts per million for solar-like oscillations to many orders of magnitudes for the most energetic phenomena in the universe like hyper-novae.

In the understanding of this broad range of variable phenomena, the ESA Gaia mission is unique thanks to its comprehensive approach that combines ultra-precise astrometric measurements to nearly simultaneous photometric, spectro-photometric and spectrometric measurements.
We summarize in Sects.~\ref{Sect:Gaia} and \ref{Sect:variableObjects} some facts of Gaia relevant to the study of variable stars. We 
illustrate in Sects.~\ref{Sect:PCA} and~\ref{Sect:classification} how Gaia's quasi-simultaneous multi-band data can benefit variability studies,
and the preparatory work done within the CU7 DPAC, in the classification of variable objects. Conclusions are drawn in Sect.~\ref{Sect:conclusions}.

\section{The Gaia observable}
\label{Sect:Gaia}
As with any multi-epoch survey, Gaia has specific properties related to its types of measurements and corresponding accuracies, and to its sampling strategy.

{\bf Types of measurements:} For the detailed description of the payload, we refer you to the Gaia webpage (http://www.rssd.esa.int/Gaia). Gaia will "see" the astronomical objects with different instruments and variability studies will benefit from all these. They are:\\
{\em 1)~The AF Astrometric Field:} The AF will provide the mean G-band photometry and G-band "epoch photometry" with its high photometric precision for stars from 6 to 20 mag. For a transit measurement, the precision is estimated to be at the level of a millimag at the bright end to 20 millimag at the faint end. Therefore, among the 1 billion observed objects, Gaia will systematically detect and describe variable objects with the amplitude of variation and the associated times-scale. The AF will also provide the astrometry. The parallax and G-band photometry combined will then give the absolute G-band magnitude. However, it is important to remember that bolometric corrections should be applied to get the luminosities.\\
{\em 2)~The BP (Blue) RP (Red) Spectrophotometer:} The mean BP and RP spectrophotometric measurements, may allow the determination of temperature, log g, metallicity and extinction. Per transit integrated BP and RP epoch photometry will allow to detect variability in different bands quasi-simultaneously. It is not yet known if full spectra can be used on a per transit level.\\
{\em 3)~The RVS Radial Velocity Spectrometer:} The RVS will give access to radial velocities and rotational velocities. The RVS instrument may also allow a multi-transit analysis with the radial velocities as well as changes in the spectra from one transit to another for the brightest stars.\\
{\em 4)~Finally, the SM Sky-Mapper:} The SM will also deliver transit photometry in the same passband as the AF measurements.

The diversity of the data is unprecedented. Gaia will not only give positions in the HR diagram of variable stars, it will also give the motion of these objects in the HR diagram, as shown for a few examples in Spano \etal\ \cite{Spanoetal2009}.

{\bf Observing strategy:} Gaia will measure positions over the entire sky 70 times on average during its 5 year mission (with a possible one year
extension). Globally, the number of measurements per object depends on the ecliptic latitude and may be up to about 250 transits per object. This is valid for the AF, SM, BP and RP instruments. We note that the RVS instrument uses 4 CCDs (in the across scan direction), instead of 7, hence there will be less measurements from the RVS.

\section{Number of variable objects and variability behaviours}
\label{Sect:variableObjects}
The number of variable sources that Gaia will be able to detect is still uncertain, which demonstrates a lack of knowledge in the domain.
We estimate it between 50 million to 150 million variables, however  these numbers depend obviously on the eventual precision reached by the satellite once in space. Space missions like Kepler and CoRoT should help to better pin down this number. The diversity of variable phenomena can be seen in the variability tree presented by Eyer and Mowlavi \cite{EyerMowlavi2008}. 
We can also regroup variability with respect to behaviour types:\\
{\bf -Periodic Objects:}
Gaia has very good performances to recover the correct period of periodic objects whose light curves can be modelled with a Fourier series with a low number of harmonics, as shown by Eyer and Mignard \cite{EyerMignard2005}.
This results from the semi-regular sampling which arises from the scanning law.
The spectral window of Gaia sampling presents high peaks at high frequencies, which avoids too numerous aliases.\\
{\bf -Semi-regular, irregular variables:}
The characterization of semi-regular and irregular variables might be difficult due to the too sparse sampling of Gaia time series.
In addition, those objects will pollute the samples of periodic objects.
The processing of semi-regular and irregular objects will thus be a challenge for the exploitation of Gaia data.\\
%
{\bf -Transient Variables:}
Transients, such as microlensing events or supernovae will be treated by the group performing the photometric reduction in Cambridge.
Some of those objects may demand a rapid response from the scientific community, for which an alert system is being put in place by this group. As shown by Mignard \cite{Mignard2010}, in the worst case the received observations might be 48 hours old. The most prompt observations could be less than 12 hours old.

\section{A solution to multi-band data: Principal Component Analysis}
\label{Sect:PCA}
One peculiarity of Gaia is its quasi-simultaneous multi-band data, integrated BP, RP, and G magnitudes, which offers a multitude of potential advantages for Gaia variability processing. The luminosity changes of most types of variable stars, be it regular or not, regardless of the time-scale of the variations, are correlated in the different wavelengths. These correlations may be exploited for many purposes: to detect variability more reliably, to distinguish microvariability from noise, to estimate periods more correctly, to study the properties of the noise, and to use derived quantities as attributes in classification. Principal component analysis (PCA) is a simple, well-developed and promising tool to achieve these improvements.

Suppose that we have $N$ observations in $M$ bands for a specific star. Disregarding the times, we can imagine this multi-band data set as an $M$-dimensional point cloud of the $N$ observations. After centring and appropriately scaling the data, the form of this cloud is characteristic: if the star is non-variable, it is near-spherical, since the only origin of variability is the measurement noise which is approximately uncorrelated over time, whereas if it is variable, correlations and variability exceeding the measurement error will imply stretching and distortion of this ball. PCA first finds the direction of the largest variability, the longest elongation of the point cloud; then the direction of the next largest variability in the subspace orthogonal to the first principal axis, and so on. The successive directions are called the principal components PC$_1$, PC$_2, \ldots$ of the data; they are related to the original data by a linear transformation.

Their use is many for variability analysis. The value of the variance of PC$_1$ and its ratio to the total variance can be used to distinguish variable sources from non-variables. Adding up the coherent variations in all bands for PC$_1$ implies a better signal-to-noise ratio, and thus more chance to find small-amplitude variables. The better signal-to-noise ratio of PC$_1$ may offer more precise period estimates. The principal components with smallest variance, PC$_{M-k}, \ldots,$ PC$_M$ yield an idea about the quality of the noise estimates. The coefficients of the original bands in PC$_1$ may help characterize the variability type.
\begin{figure}[t]
\centering {
\includegraphics[scale=0.52]{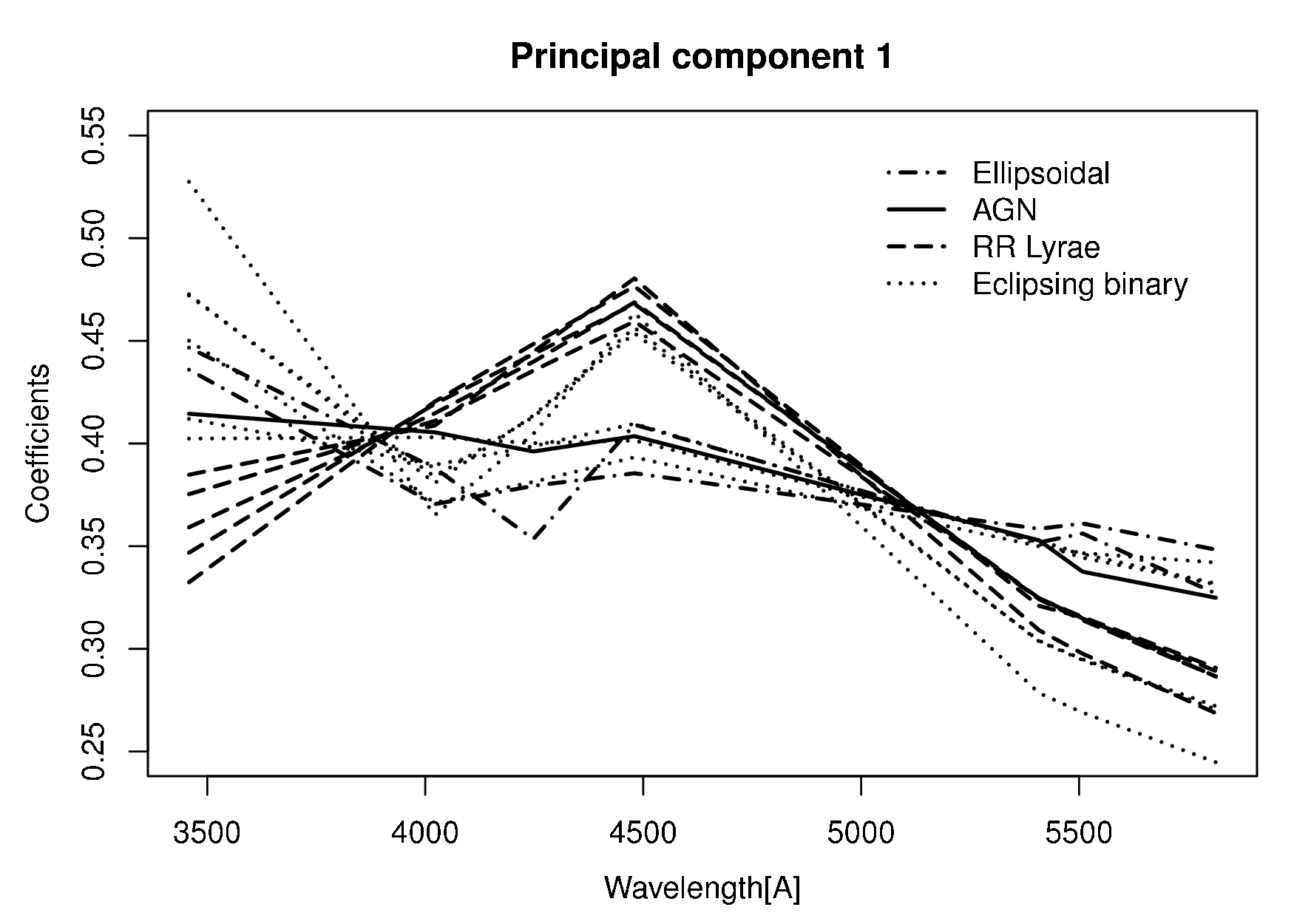}}
\caption{\small \label{fig_nonscaledPC1}  First principal component PC$_1$ for different star types as functions of the mean wavelength of the Geneva photometric bands. AGN (3C273) has variations having a small trend of being larger when bluer, ellipsoidal variables as well. Pulsating stars like RR Lyrae stars have the largest amplitude in B2 Geneva band, and eclipsing binaries have diverse behaviours.}
\vspace{-0.5cm}
\end{figure}

Investigations so far were performed on 150 non-variable stars from Geneva photometry and on 2000 non-variable and 2000 variable stars from SDSS Stripe 82 (\cite{Sesar07}). The analysis of Geneva photometry stars confirmed the ability of PCA to detect small-amplitude variability missed by conventional single-band analysis and the advantages of the period search performed on PC1 (\cite{Bartholdi}).  Moreover, Figure \ref{fig_nonscaledPC1}, showing the PC$_1$ coefficients for several objects from Geneva photometry, suggests it as an aid in distinguishing between variability types. The research on Stripe 82 directed attention to many problems that present a challenge in analyzing Gaia data:  finding efficient ways to deal with the sharply different errors in the different bands, to filter out outliers and erroneous data and cope with wrong error estimates. The results suggest that PCA combined with other, robust statistical tools can indeed help variability detection, characterization and classification, and thus provide valuable addition to the wealth of informations furnished by Gaia.

\section{A test: Classification of Hipparcos variable sources}
\label{Sect:classification}
The best way to get ready for classifying the Gaia variable sources is probably to gather experience with available data from other surveys. Learning how to deal with difficulties related to different survey idiosyncrasies can help getting prepared for some possibly difficult and unexpected Gaia features. In this context, the Hipparcos mission is an obvious starting point. It includes many of the brightest stars in the sky, which are amongst the best known objects, and the number of measurements, the total mission duration and time sampling are somewhat similar. In addition, a systematic automated classification of the Hipparcos variable sources has not been achieved yet.

The {\em variability type} of each of 2712 Hipparcos periodic variables is provided by a survey of recent literature or by private communications from DPAC CU7 scientists who maintain lists of variables.

A number of {\em attributes} are computed to characterise the sources.  Some of them reflect the global properties of the stars, such as the mean color or the absolute brightness, others describe some of the Hipparcos light-curve features. A number of statistical parameters are derived from the magnitude distribution, a period search is carried out and the folded light-curves are modelled with Fourier series. Many harmonics are fitted, but only those that are significant according to an F-test are kept. The number of harmonics is also limited if there are gaps in the time sampling to avoid non-physical large model excursions in regions devoid of measurements.

A major difficulty is to evaluate which of the many computed attributes is really useful in the classification process, also knowing that many of them are highly correlated. This is where the {\em Random Forest} methodology comes to the rescue as it includes an internal measure of the attribute importance (Breiman, \cite{Breiman2001}). A ranked list of ``not-too-correlated'' attributes can be built iteratively. The most important one is selected and all other highly correlated attributes are discarded. The second most important is added to the list and again, highly correlated attributes are removed, and so on.

An algorithm proposed by Svetnik \etal\ \cite{Svetniketal2004} can then be used to determine the minimum list of attributes. In summary, the
classification errors are first derived through a cross-validation approach with the top two most-important attributes. When new attributes are added in order of decreasing importance from our ranked list, the classification error diminishes, until it reaches a minimum value ``plateau''. The minimum attribute list is the one used when the plateau value is first reached.

The classification results obtained with the Random Forest method are slightly better than those obtained with the previously used Bayesian Network. There is some confusion within groups of related stars, such as different eclipsing binary types or single and multiple mode Cepheids and W Virginis types. About 20 EBs (10\%) get scattered into other types, while about 30 stars from different types are believed to be EBs. There is also some confusion between EW and Ellipsoidal stars, but to a lesser extent. SPB and Be stars are also mixed up. Apart form these cases, and a handful of exceptions, the type of all other stars can be perfectly predicted. One should keep in mind that these excellent results are obtained with a sample of well-known and well-behaved Hipparcos stars. This sample provides, however, a solid basis for the training set required to continue our work on other surveys.

\section{Conclusions}
\label{Sect:conclusions}
Gaia will reveal the variable universe in an unprecedented way by providing fundamental properties for each object and a description of its variability behaviour, and allowing the description of the properties of variable stars groups. The search for peculiar objects/behaviours will be possible in a dataset of one billion stars. Gaia will also help and be complementary to other projects having different sampling strategies/accuracies (such as Kepler, OGLE, LSST, etc\ldots) thanks to astrometric and complementary information. While Gaia will solve many problems, it will also raise many interesting questions. Thanks to the bright limit of the photometric range covered by Gaia, follow-up campaigns and further scientific studies will be possible with small telescopes.

With the coming monumental dataset, new efficient methods should be developed for looking at the data, analysing it, browsing and searching objects/groups of objects. The work to be done to maximize the mission scientific return pertains to different fields: mathematics, statistics, simulations, computer sciences and data handling. Many of these aspects have been tested with Gaia simulations and real data before getting the actual Gaia data. Part of this work is achieved within CU7 DPAC, whose goal is to produce the ESA Gaia intermediate and final catalogues of variable objects for release to the scientific community.


\begin{thebibliography}{99}
\bibitem[Bartholdi, 2005]{Bartholdi} Bartholdi, P. 2005, Gaia document: VSWG-PB-001
\bibitem[2001]{Breiman2001} Breiman L. 2001, Machine Learning, 45(1), 5
\bibitem[2005]{EyerMignard2005} Eyer, L., Mignard, F. 2005, MNRAS, 361, 1136
\bibitem[2008]{EyerMowlavi2008} Eyer, L., Mowlavi, N. 2008, J. Phys. Conf. Ser., 118, 2010
\bibitem[Hastie \etal, 2009]{Hastie} Hastie, T., Tibshirani, R., Friedman, J. 2009,  The Elements of Statistical Learning, Springer, ISBN 978-0-387-84857-0
\bibitem[2010]{Mignard2010} Mignard, F. 2010, June 23-25 Science Alert Workshop presentation, Cambridge, UK
\bibitem[Sesar \etal, 2007]{Sesar07} Sesar, B., \etal\ 2007, AJ, 134, 2236
\bibitem[2009]{Spanoetal2009} Spano, M., Mowlavi, N., Eyer, L., Burki, G. 2009, AIP Conf. Proc., 1170, 324
\bibitem[2004]{Svetniketal2004} Svetnik V., \etal\ 2004, in: Roli F, Kittler J, Windeatt T (eds) Lecture notes in computer science, Springer, 3077, 334
\end{thebibliography}
\end{document}